\newcommand{\gsim}{\;\raisebox{-.3em}{$\rlap{\raisebox{.45em}{$>$}}\!\!\sim$}\;}
\documentstyle[12pt, epsfig] {article}
\begin{document}
\begin{center}
{\large \bf On The Phase Transition in D=3 Yang-Mills Chern-Simons
Gauge Theory}\end{center}
\begin{center}
John M. Cornwall*
\end{center}
\begin{center}
{\it Physics Department, University of California at Los Angeles\\
405 S. Hilgard Ave., Los Angeles Ca 90095-1547}
\end{center} 
\begin{center}
{\bf Abstract}
\end{center}

$SU(N)$ Yang-Mills theory in three dimensions, with a Chern-Simons term
of level $k$ (an integer) added, has two dimensionful coupling constants,
$g^2 k$ and $g^2 N$; its possible phases depend on the size of $k$ relative
to $N$.  For $k \gg N$, this theory approaches topological Chern-Simons
theory with no Yang-Mills term, and expectation values of multiple
Wilson loops yield Jones polynomials, as Witten has shown; it can be
treated semiclassically.  For $k=0$, the theory is badly infrared
singular in perturbation theory, a non-perturbative mass and subsequent
quantum solitons are generated, and Wilson loops show an area law.
We argue that there is a phase transition between these two behaviors at
a critical value of $k$, called $k_c$, with $k_c/N \approx 2 \pm .7$.
Three lines of evidence are given:  First, a gauge-invariant one-loop
calculation shows that the perturbative theory has tachyonic problems if
$k \leq 29N/12$.  
The theory becomes sensible only if there is an additional dynamic source
of gauge-boson mass, just as in the $k=0$ case. Second, we study in a rough approximation the
free energy and show that for $k \leq k_c$ there is a non-trivial vacuum
condensate driven by soliton entropy and driving a gauge-boson dynamical mass
$M$, while both the condensate and $M$ vanish for $k \geq k_c$. Third, we study possible
quantum solitons stemming from an effective action having both a 
Chern-Simons mass $m$ and a (gauge-invariant) dynamical mass $M$.  We show
that if $M  \gsim  0.5 m$, there are finite-action
quantum sphalerons, while none survive in the classical limit $M=0$, as shown
earlier by D'Hoker and Vinet.  There are also quantum topological vortices
smoothly vanishing as $M \rightarrow 0$.\\[.2in]
UCLA/96/TEP/7\\
February 1996\\ 
\footnoterule
\noindent *Electronic address:  cornwall@physics.ucla.edu
\newpage
\section{Introduction}
We study here a three-dimensional Euclidean Yang-Mills theory with a 
Chern-Simons mass term added\cite{DJT}, or just YMCS theory for short.
As is well-known, the Chern-Simons term generates a gauge-boson mass
$m$ of magnitude $k g^2/4 \pi$, where $k$ is an integer (which we can
and will choose to be non-negative).  Although a massless pole still
remains in the gauge-boson propagator, this pole is associated with
kinematical factors which remove all infrared divergences, at least in
gauge-invariant quantities.  It may therefore seem, at a naive glance, that
YMCS has a  perfectly respectable perturbation theory, at least in the sense
that every term in the series is well-defined, even if the series might not
converge.  Of course, one might expect perturbation theory to fail for small
$k$, because (see below) one expansion parameter of YMCS is $N/k$ (modulo a numerical
factor) and becomes large for $k \ll N$.  But there is more to it than just
the size of this parameter; there is a critical minus sign which is 
analogous to the sign of the $\beta$-function in $d=4$ gauge theory, which
drives the phenomena we discuss.  In the eighties a number of authors calculated some one-loop
terms in the perturbation series\cite{DJT,PR} for YMCS, especially for the conventionally-defined Feynman propagator.  Unfortunately, because of the gauge dependence
of the conventionally-defined self-energy, it was not possible to see
this sign structure gauge-invariantly from these
calculations.  But we will show in a gauge-invariant way that YMCS theory
has a tachyonic problem in physical amplitudes, in spite of the fact that the Euclidean
Chern-Simons term has a factor of $i$ which makes the mass $m$ non-tachyonic.
This tachyonic problem occurs for the same reason (gluon spin couplings) that there is an infrared renormalon pole
in the running charge in $d=4$ gauge theory, as we will elaborate below.
In $d=4$ one is used to associating this tachyonic pole with asymptotic
freedom, and we will use this same phrase as a shorthand for the sign
structure of YMCS, even though as a $d=3$ theory it is superrenormalizable
and has no renormalization group.

The essential point is that when $k$ is less than a critical value $k_c$,
of $O(N)$, the mass $m$ is too small to overcome the tachyonic tendencies
associated with what would be called asymptotic freedom in $d=4$.  Then
the situation is analogous to the $k = 0$ case, which is clearly 
non-perturbative\cite{L,GPY,C,CHK}.  The cited authors discuss the need for
generation of a non-perturbative dynamical gauge mass $M$ of $O(Ng^2)$, to
cure the infrared divergences of perturbation theory.  Dynamical mass generation
is driven by (and drives) the formation of a gauge-boson condensate,
as reflected in a positive value of $\langle (G^a_{ij})^2 \rangle$.  In spite of being
$O(g^2)$, the dynamical mass vanishes to all orders of perturbation
theory, since there is no acceptable mass counterterm to put in the
action.  Although the local gauge symmetry is exactly preserved in the
course of dynamical mass generation, the massive gauge bosons are
necessarily accompanied by massless excitations which, like Goldstone bosons
in symmetry breaking, are ``eaten" and do not appear in the $S$-matrix
as massless poles.  They do have profound physical effects, however.  The
massless excitations correspond to long-range pure-gauge parts of the gauge
potential, which have topological significance and generate, among other
things, the area law for Wilson loops\cite{C79,C95,CY}.  The confinement
(or area-law) mechanism, just as in lattice gauge theory\cite{T93}, is one
of topological linkage of a condensate of closed vortices (which can be cut open by the
formation of a monopole-antimonopole pair) with the Wilson loop.  In fact,
the area law is described via the standard Gauss linking integral for 
two closed strings.  Moreover, the fluctuations of the Chern-Simons term are
also described\cite{C95,CY,TV} as averages over powers of the linking number,
in this case of vortex strings with themselves or other vortex strings in
the vacuum.  These vortices are one of several types of solitons which 
occur in the dynamically-massive theory (another is the sphaleron 
corresponding to no Higgs symmetry breaking\cite{C77,CT}) but not in the
classical massless theory; that is, these solitons owe their existence to
a dynamical mass.  Since this mass is purely a non-perturbative quantum
effect, we call these quantum solitons.  The entropy of the vortex solitons
is larger than their free energy, and a condensate of the vortices forms,
which is responsible for dynamical mass generation (for example, Lavelle
\cite{La} has shown that the dynamical gauge-boson mass behaves rather
like a constituent quark mass, with the squared mass vanishing at large momentum $p$ at a rate
involving the relevant condensate expectation value:
$\langle g^2 (G^a_{ij})^2 \rangle /p^2$).  

We will show that in the opposite case $k > k_c$, the tachyonic problem and its cures, namely generation of a 
dynamical mass and condensate, go away and the theory smoothly merges,
as $k \rightarrow \infty$, into Witten's\cite{W,H} topological theory
which is accessible by semiclassical means.  Because the dynamical mass and condensate
vanish, there are no quantum solitons, and it is known that there are no classical
solitons of finite action in YMCS theory\cite{DV}.

The purpose of the present work is to describe the nature of this unusual
phase transition and to make some (probably not very accurate) estimates
of some of the numbers involved, such as the critical value $k_c$ of $k$.
The essence of this phase transition is the analog\cite{C,CHK,C89} of asymptotic freedom in four dimensions, as indicated by a crucial sign coming from the
spin-dependent gauge-boson couplings\cite{NH}.  Of course, there is no
renormalization group in $d=3$ gauge theory, so the usual calculation of
a $\beta$-function which reveals this sign structure in $d=4$ cannot be done.
But it is possible to calculate a running charge\cite{C,CHK,C89} using the
pinch technique, which identifies\cite{C,CP} a gauge-invariant gauge-boson 
proper self-energy from the $S$-matrix, and the square of this running charge 
becomes negative for sufficiently small momentum, because of a tachyonic (real Euclidean) zero in a gauge-invariant proper self-energy. If this zero
were missing, many or all of the effects we find would not occur.  

We can give a simple quantum-mechanical analog to illustrate this point
about signs.  Consider the two forms of the quartic oscillator Hamiltonian:

\begin{equation} H = \frac{1}{2} p^2 \pm \frac{1}{2} \omega^2 x^2 +
\frac{\lambda}{4} x^4 \end{equation}                                      

The dimensionless expansion parameter is $\lambda / \omega^3$, and can
be large or small.  For the positive sign, however, the qualitative
behavior is not changed as this parameter goes from small to large; to be
sure, perturbation theory does not converge (for any value), but it is
Borel-summable, and any number of techniques (e.g., variational) serve to
estimate the energy levels, etc, with acceptable accuracy.  But for the
negative sign the potential has two wells, Borel summation fails, and the
system behavior is quite different for large and small values of
$\lambda / \omega^3$.  For large values there is a high barrier separating
the two wells and these communicate with each other only by exponentially-
suppressed tunnelling, but for small values the barrier height $\lambda
/ \omega^2$ is less than the perturbative energy scale $\omega$, and
the hump separating the two wells is only a small perturbation.  In YMCS
gauge theory the expansion parameter is $Ng^2/m$ (more generally,
$Ng^2/p$ at momentum scale $p$), or equivalently $N/k$, and the negative
sign in the quartic oscillator potential (1) corresponds to asymptotic
freedom.  By analogy, one expects different phases as $k$ varies relative
to $N$, as we shall argue for on other grounds.  Correspondingly, if the
dynamics of YMCS were not asymptotically-free, one might expect quite
different behavior, with no essential difference between large and small $k$.
(Ref. \cite{C89} has some remarks on what $d=3$ YM theory would look like
in this case).

We now describe the three approaches taken in this paper:
\begin{enumerate}
\item {\bf Gauge-Invariant One-Loop Self-Energy.}  For details see Section 2.  One can define from the
$S$-matrix for any process propagators and vertices whose proper parts are
completely independent of the chosen gauge, and which satisfy ghost-free
Ward identities\cite{C,CP}.  These were previously calculated\cite{C,CP,
CHK,C89} at one-loop order for $d=3,4$ YM theory.  In the present work we
calculate the gauge-invariant proper self-energy for YMCS theory, extending the work of
Pisarski and Rao\cite{PR}, which is a calculation of the conventional
self-energy in the Landau gauge.  Our extension is a straightforward if 
lengthy piece of work.  We quote here some useful results.  For YM theory
with no CS term, the inverse propagator with gauge-invariant self-energy, $\hat{\Delta}_{ij}^{-1}(p)$, is:
\begin{eqnarray} \hat{\Delta}_{ij}^{-1} & = & (p^2\delta_{ij} - p_i p_j) d(p)
+  p_i p_j/\xi \\
d(p) & = & 1 - \pi b g^2/p  \end{eqnarray}                              
where the {\it gauge-invariant} constant $b$ is:
\begin{equation} b = 15N/32\pi \end{equation}                                
and $p$ is the magnitude of the Euclidean three-momentum; $\xi$ is an
arbitrary gauge parameter.  There is a tachyonic zero at $p = \pi b g^2$,
because of the minus sign in (3); this zero leads to unphysical effects such
as an imaginary running charge, as we describe later.  

For YMCS theory at level $k$, there is a tree-level mass $m$:              
\begin{equation} m = k g^2/4 \pi \end{equation}
and the propagator $\hat{\Delta}_{ij}^{-1}$ has two terms, an even term with
the kinematics of equation (2) and an odd term, to be given later.  The
even term we write as, with the notation of (2),
\begin{equation} d(p) = 1 - \hat{A}(p), \end{equation}                    
with $\hat{A}$ a complicated function given in Section 2.  At $p=0$ this
function has the value
\begin{equation} \hat{A}(0) = 29Ng^2/48 \pi m = 29N/12k \end{equation}   
and $d(0)$ is negative (tachyonic) if
\begin{equation} k \leq k_c = 29N/12. \end{equation}                                   Since $\hat{A}(p)$ is a positive monotone decreasing function of $p$
(behaving like $1/p$ at large $p$), there will always be a tachyonic zero
in $d(p)$ for some $p$ if $k \leq k_c$ as given in (8).

One may anticipate that this tachyonic zero in $d$ will be removed\cite{C,CHK,C89} by the
same mechanism in YMCS that operates for just YM theory with no CS term:  A dynamical mass $M$ is generated by condensate
formation, in the case of YM theory replacing  the $1/p$ in $d$ of equation (3) by something like $1/(p^2 + 4 M^2)^{1/2}$, with $M$ large enough to keep
$d$ positive at zero (and thus at any real Euclidean) momentum.  Similarly,
$m$ in equation (7) should be replaced by something like $(m^2 + M^2)^{1/2}$
with $M$ large enough to keep $d$ positive.  We now discuss how this might
happen.

\item {\bf Mass and Condensate Generation At Small k.}  These results are
detailed in Section 3; we believe they are qualitatively 
correct although perhaps far from quantitative.  Define the partition
function $Z$ and its logarithm as usual:

\begin{eqnarray} Z(k) & = & \int (dA) \exp(-S_{YMCS}) \\    
 Z(k) & = & \exp(- \int d^3x \epsilon) \end{eqnarray}
It is easy to show that $\epsilon$ is real and that $\epsilon(k \neq 0)$ is always larger than
$\epsilon(k=0)$; for small $k$, the fractional increase in $\epsilon$ is
$O(k^2/N^2)$.

This increase in $\epsilon$ is in the direction to disfavor condensate
formation, since condensate entropy tends to lower $\epsilon$.  To 
understand this effect of non-zero $k$ it is first necessary to review
earlier work on condensate formation at $k=0$.

Previously\cite{C94} we have given the exact form
of the action for pure YM ($k=0$) theory in its dependence on the zero-momentum
matrix elements of the condensate operator
 \begin{equation} \theta = \frac{1}{4}(G^a_{ij})^2    \end{equation}   
and shown that $\epsilon$ (or equivalently the spatial density of
$\beta F$, where $F$ is the free energy, in thermal field theory) has a
minimum for positive $\langle \theta \rangle$, and the minimum value is
negative.  This shows that a condensate has formed, with condensate entropy
outweighing the positive internal energy.  We have also given arguments
\cite{C89,C94} consistent with Lavelle's\cite{La} work that condensate formation drives the generation of a
dynamical mass $M$, with $M$ depending on $\theta$ as $(g^2 \theta)^{1/4}$,
as naive dimensional reasoning predicts.  A simple crude model of the free
energy $\epsilon$ was constructed\cite{C89,C94}, based on adding the mass
$M$ by hand to the gauge-invariant propagator $\hat{\Delta}$ discussed
above, and the resulting form for the free energy was consistent (in a non-
trivial way!)  with the required\cite{C94} dependence on $\theta$.  The
addition of mass to this propagator was justified by an earlier\cite{CHK}
investigation of a non-linear gauge-invariant Schwinger-Dyson equation
for $\hat{\Delta}$, which showed that, because of the sign of the $b$ term
in equation (3), this equation required a dynamical mass.  The value of the
mass could not be predicted, but could be bounded below; the bound is
roughly $2bg^2$ in terms of $b$ of equation (4).

In the present work we extend these considerations to YMCS, in a crude way.
To be more exact would require once again consideration of the non-linear
Schwinger-Dyson equation for the gauge-invariant propagator, which we have
not yet attempted.  The same general effects are operative; because of the
sign structure of the one-loop perturbative propagator, would-be tachyonic
effects require more than just the CS mass $m$ for their cure when $k$ is
small.  It is plausible that the outcome of the Schwinger-Dyson equation, if
really solved, would be to replace the perturbative YMCS mass $m$ by a nonperturbative value of $O((m^2 + M^2)^{1/2})$, where now M stand for the
contribution of the condensate to the mass; that is, $M \sim (g^2\theta)^{1/4}$ as before.  Now we simply take the one-loop YMCS propagator with this
replacement for the mass, and repeat the earlier\cite{C89,C94} work on pure
YM theory.  The result is consistent with the general remarks made above,
in that the lowest-order correction is $O(k^2/N^2)$ and increases the free
energy.  As $k$ increases a critical value $k_c$ is reached where the free energy
is positive and the condensate and its associated dynamical mass vanish.
Because our construction of the free energy is based on a one-dressed-loop
approximation we find the same value of $k_c$ as in one-loop
perturbation theory (see (8)).  We also find a critical exponent for
dynamical mass generation:

\begin{equation} M \sim (k_c - k)^{1/2}\;for\;k<k_c   \end{equation}

\item {\bf Quantum and Classical Solitons.}  More evidence for the two-phase
structure of YMCS theory is found by looking for classical and quantum
solitons.  In the classical theory, with no dynamical mass $M$, this has
already been done by D'Hoker and Vinet\cite{DV}.  The results are that there
are no Euclidean classical solitons with finite action, either of the
Abelian vortex type or of the sphaleron (spherically-symmetric) type.  In
the latter case, a peculiar solution exists in which the soliton field has an
accumulation point of zeroes at the origin; its action is infinite.  The absence of classical
solitons  with finite action is consistent with the idea that the $k>k_c$ phase has no
condensate and can be treated semiclassically.

In the other phase, with a dynamical mass $M$, these solitons are
profoundly modified, and one can find quantum solitons of both the vortex
type and the sphaleron type.\footnote{The author and B. Yan have made a
preliminary study of the analogous solitons in Euclidean YMCS theory with a
{\it real} Chern-Simons coefficient.  The CS coefficient then serves as a Lagrange
multiplier for specifying the expectation value of the CS action.  The resulting solitons are real, and the vortex is twisted; this twist contributes to the
CS term\cite{C95,CY,TV}.  Results will be reported elsewhere.}  These are found as classical solutions of
an effective action, containing not only the usual YM and CS terms but also
\cite{C74} a gauge-invariant mass term for the dynamical mass; this is
just a gauged non-linear sigma model.  It is well-known that this added
mass term does not lead to a perturbatively-renormalizable theory, because
of divergences associated with the implicit assumption that the mass $M$
is a ``hard" mass, surviving at large momentum.  In fact, the dynamical
mass is a soft mass, vanishing\cite{CH86,La} like $p^{-2}$ (modulo logarithms) at large
momentum $p$.  Since we should not and will not use the effective action
beyond the classical level, we will treat the mass $M$ as a constant, 
although this introduces a spurious logarithmic divergence in the action
associated with the gauged non-linear sigma model.  The true action,
with all quantum corrections, is finite.

With the dynamical mass added, there are actually two different 
propagator-pole masses, both non-zero; as $M$ approaches zero, the
heavier mass becomes the CS mass $m$ and the lighter one approaches zero
(which it will be recalled is also a propagator mass of classical YMCS theory).
At $m=0$, both masses merge into the dynamical mass $M$.  In addition to these
poles, there is also a zero-mass excitation corresponding roughly to a
Goldstone boson (although there is no symmetry breaking), but this is better
identified as a long-range pure-gauge excitation of the gauge potential
which yields such effects as\cite{C79,CY} confinement and a string tension.

The effective action is complex because of the CS term, which has an $i$
factor in Euclidean space.  As a result, the general soliton solution of the effective action is also complex. In view of the fact that the partition function
is real, there can be solitons which have complex action but which occur
in complex-conjugate pairs (equivalently paired under $k \rightarrow -k$), or complex solitons which have real action.  We
have only found the latter, both for the vortex and for the sphaleron.
These two solitons have different fates as the dynamical mass $M$ goes to
zero, that is, as $k \rightarrow k_c$.  The vortex is Abelian, and is an
extension of the well-known Nielsen-Olesen vortex; its gauge potentials can be explicitly written in
terms of Hankel functions of imaginary argument.  This vortex involves both of the
pole masses mentioned above in such a way that it has finite YMCS action
by virtue of a cancellation between terms involving these different masses,
and smoothly vanishes as $M$ goes to zero.  The sphaleron soliton is rather
different; it can only be found numerically.  We have found numerical
solutions for small values of $m/M$, which will be displayed in Section 4.
As this parameter increases it becomes increasingly harder to find solutions to
the equations of motion.  We have used a simple variational approach 
for larger values of $m/M$ which suggests that the sphaleron soliton becomes
singular at $m/M \gsim 0.5$; the singularity is of the same general type
as found by D'Hoker and Vinet\cite{DV}.  

In both cases, the solitons we have found decouple from the theory when
$m/M$ or equivalently $k/N$ is large enough, which is consistent with the
idea that YMCS theory becomes semiclassical at large $k$.

To make further progress would require an accurate evaluation of the
contribution of the solitons, including their entropy, to the partition function.
This is difficult and uncertain, and we have not attempted it.
\end{enumerate}

It appears that YMCS theory, while not possessing any immediately obvious
applications to particle physics, is an interesting testing ground for
various non-perturbative phenomena of field theory.  One check of the ideas
presented here might be through lattice-gauge simulations.  Other directions
worthy of investigation are to search for an approximate duality between
$k$ and $N$, in the spirit of Seiberg-Witten\cite{SW} duality, or to extend
the theory to supersymmetric YMCS, in light of the fact that chiral fermions
can induce a CS term\cite{SUSY}.

\section{Gauge-Invariant One-Loop Perturbation Theory}

In this Section we will calculate a {\it gauge-invariant} one-loop 
self-energy for YMCS theory, using the pinch technique\cite{C,CP}.
The pinch technique adds to the conventional self-energy some new terms
defined by the $S$-matrix; these new terms, among other things, cancel
the dependence of the conventional self-energy on the choice of gauge
\footnote{It has been remarked\cite{C89} that at least at one-loop
level the pinch technique gives the same result as the Feynman-gauge
background field technique.}.
Since the usual self-energy has already been calculated by Pisarski and
Rao\cite{PR} in the Landau gauge, we need only compute the extra terms,
also in the Landau gauge.  The result is independent of the gauge, as one
may readily check by adding the appropriate gauge terms to both the 
Pisarski-Rao terms and the terms we find here.

To understand the pinch technique, consider the one-loop graphs for the
$S$-matrix element of two-particle scattering, where the external lines can have different masses and spins, be
in arbitrary representations of $SU(N)$, etc.  The only requirement is that
they be on-shell.  All these graphs for fermions (except for external-line wave-function
renormalization) are shown in Fig. 1a, b, c, d, f, g, i.  (For the moment
ignore the graphs with heavy vertices.)
The conventional propagator comes
from Fig. 1a, b, c with, of course, no external lines attached.  But just
the sum of these three graphs is not gauge-invariant; all the rest of the graphs must be
added to get a gauge-invariant $S$-matrix.  The pinch technique identifies parts of graphs d, f, g, and i
which act exactly like propagator parts and which, when added to the usual
terms, yield a gauge-invariant result.  This must happen because all terms
in the $S$-matrix with the kinematic structure of propagator exchange between
bare vertices have different dependence on the kinematic variables than any
other set of terms (e. g., they are independent of external-line masses and
of energy variables, except for trivial external-line wave functions).

Consider now the vertex labeled {\bf i} in Fig. 1d, associated with a factor
$\gamma_i$.  There will be a term 
$\sim k_i$  multiplying this vertex coming from gauge-boson propagator
parts or from three-boson vertices.  This triggers the Ward identity
\begin{equation} k_i \gamma_i = S^{-1}(p) - S^{-1}(p-k) \end{equation}
where $S(p)$ is the external-line propagator of momentum $p$.  But $p$ is
on-shell, so $S^{-1}(p) = 0$, and the other term in (13) cancels out the
propagator of momentum $p - k$.  The result is a graph with the structure
shown in Fig. 1e, where the heavy vertex indicates a pinch has taken place.
Similarly, in any but the Feynman gauge the box graphs Fig. 1f, g and the
vertex graph Fig. 1i have pinch parts, as shown in Fig. 1h, j.  
 
We will next report on the calculation of the pinch graphs (Fig. 1d, h, j),
which is straightforward but somewhat lengthy.  One comment is needed about
the pinch in graph Fig. 1i.  The group-theoretic factor of this vertex is
$C_R - N/2$, where $C_R$ is the Casimir invariant for the external lines in
representation $R$ of $SU(N)$, and the $N/2$ is half the adjoint Casimir.
The $C_R$ part cancels the gauge dependence in external-line wave-function
graphs (not shown), which occur with weight 1/2 twice for each line.  The
only pinch cancellation relevant to the propagator comes from the $N/2$ part,
and therefore we define the pinch graph Fig. 1j to have the group-theoretic
factor $-N/2$.  Note that this graph vanishes, by dimensional regularization,
in the $m=0$ (pure YM) theory, but not in YMCS theory.  Also note that graphs
Fig. 1e, j must be multiplied by two because of the two external lines.

First we establish some notation.  Define the scalar one-loop integral
with two masses:
\begin{equation} J(m_1, m_2; p) = \frac{1}{(2 \pi)^3} \int d^3q  
\frac{1}{(q^2 + m_1^2)((q + p)^2 + m_2^2)} \end{equation}
so that $J(0,0;p)=1/8p$ and
\begin{equation} J(0,m;p) \equiv J_1 = \frac{1}{16p} + \frac{1}{8 \pi p}
\arctan (\frac{p^2 - m^2}{2pm}) \end{equation}
\begin{equation} J(m,m;p) \equiv J_2 = \frac{1}{4 \pi p}
\arctan (\frac{p}{2m}) \end{equation}
We use the Pisarski-Rao\cite{PR} bare propagator and vertex:
\begin{equation} \Delta_0(p)_{ij} = (\delta_{ij} - p_ip_j/p^2 -m \epsilon_{ija}
p_a/p^2) \frac{1}{p^2 + m^2} + \xi p_ip_j/p^4 \end{equation}
\begin{equation} \Gamma_{ijk}(p,q,-p-q) = \delta_{jk}(2q + p)_i
- \delta_{ik}(2p + q)_j + \delta_{ij}(p-q)_k + m \epsilon_{ijk} \end{equation}
(The ghost propagator and vertex and the four-point vertex are the same as
in pure YM theory).  In (17), $\xi$ is a gauge parameter.  

The gauge-invariant propagator inverse is:

\begin{equation} \hat{\Delta}^{-1}(p)_{ij} = \Delta^{-1}_0(p)_{ij} -
\hat{\Pi}(p)_{ij}  \end{equation}
where the bare inverse is
\begin{equation} \hat{\Delta}_0^{-1}(p)_{ij} = (p^2\delta_{ij} - p_ip_j)
+ m\epsilon_{ija}p_a + p_ip_j/\xi \end{equation}
and the self-energy has the conserved\footnote{Longitudinal terms cannot
contribute to the $S$-matrix of Fig. 1.} form:
\begin{equation} \hat{\Pi}(p)_{ij} = (p^2 \delta_{ij} - p_ip_j)\hat{A}(p)
+ m\epsilon_{ija} p_a \hat{B}(p). \end{equation}
The propagator itself is:
\begin{eqnarray} \hat{\Delta}(p)_{ij} & = & (\delta_{ij} - p_ip_j/p^2) 
\frac{1}{(1 - \hat{A})(p^2 + m_R^2)}\\ & & - m_R \epsilon_{ija}p_a 
\frac{1}{p^2(1 - \hat{A})(p^2 + m_R^2)} + \xi p_ip_j/p^4 \nonumber\end{eqnarray}
and the renormalized running mass $m_R$ is:
\begin{equation} m_R(p) = m(\frac{1 - \hat{B}}{1 - \hat{A}}) \end{equation}

  Just as one does for QED, one can define running charges (in this case, two
of them, one for the parity-even exchange and one for the parity-odd) via
\begin{equation} g^2\hat{\Delta}(p)_{ij} = \delta_{ij} \frac{g_R^2(p)}
{p^2 + m_R^2} + \cdots \end{equation}
(We do not write the parity-odd term explicitly.)  Clearly, 
\begin{equation} g_R^2 = \frac{g^2}{1 - \hat{A}} \end{equation}
What we find in one-loop perturbation theory is that for $k<29N/12$,
there is a real value of the momentum $p$ for which $1 - \hat{A}$ vanishes,
and this quantity is negative for smaller $p$.  Evidently, this leads to tachyonic poles
in both the running charge and in the running mass (but not necessarily
in the propagator itself unless $1 - \hat{B}$ vanishes at the same
momentum); for momenta smaller than the pole momentum, the running charge
is imaginary.  Such behavior is physically unacceptable, and calls for
dynamical mass generation as discussed in Section 3.

Here are the one-loop results, beginning with the Pisarski-Rao calculation.
Letters in parentheses refer to the appropriate graphs of Fig. 1.

\begin{eqnarray} \hat{A}(a, b, c) & =& \frac{Ng^2}{32\pi m}\{5 + 11m^2/p^2
- \frac{\pi}{2m^3p^3}[m^2(2p^4 + 13p^2m^2/2\\ & & + 7m^4/2)                           - (p^2 - 7m^2)
(p^2 + m^2)^2(-1/2 + 8pJ_1)\nonumber\\ & & - 4p(p^4 - 13p^2m^2 + 4m^4)
(p^2 + 4m^2)J_2 ] \}\nonumber \end{eqnarray}
\begin{eqnarray} \hat{B}(a, b, c) & = & -\frac{Ng^2}{16 \pi m}
\{2 + m^2/p^2 + \frac{\pi}{4m^3p^3}[m^2(p^4 + p^2m^2 - m^4)\\
& & + (3p^2 - m^2)(p^2 + m^2)^2(16pJ_1 - 1)\nonumber\\
& & - 24p^3(p^2 - 2m^2)(p^2 + 4m^2)J_2]\}\nonumber \end{eqnarray}
Next, the pinch contributions, listed for separately for graph e and for
graphs h + j:
\begin{eqnarray} \hat{A}(e) & = & Ng^2\{9m/16 \pi p^2 + p/8m^2
-7/16\pi m + 3J_2\\
& & + J_1[4p^2(m^2 - p^2) - (p^2 + m^2)(p^2 + 5m^2)]/4p^2m^2
-2(p^2 - m^2)Q\} \nonumber \end{eqnarray} 
where
\begin{eqnarray} Q & = & -1/32 \pi mp^2 + (p^2 + m^2)^2J_1/8p^2m^4\\
& & -(p^2 + 4m^2)J_2/16m^4 - p/128m^4 \nonumber \end{eqnarray}
\begin{eqnarray} \hat{B}(e) & = & Ng^2\{-p^3/8m^4 -15/16\pi m
+ m/16\pi p^2 \\
& & + J_1(7/2 + 15p^2/4m^2 - m^2/4p^2 + 2p^4/m^4)\nonumber\\
& & + J_2(3 - 15p^2/4m^2 + p^4/m^4) - 4p^2Q\nonumber \end{eqnarray}
\begin{equation} \hat{A}(h+j) = Ng^2\{1/4\pi m - (p^2 - m^2)/m^2
(J_1 - J_2) + (p^2 - m^2)Q\} \end{equation}
\begin{eqnarray} \hat{B}(h + j) & = & Ng^2\{1/4\pi m + p(p^2 - m^2)/16m^4
\\ & & + p^2(p^2 + m^2)(J_2 - J_1)/2m^4 + 2p^2Q \nonumber \end{eqnarray}

The functions $\hat{A}, m\hat{B}$ have no infrared singularities at $m=0$
for finite $p$, or at $p=0$ for finite $m$.  So a perturbation series in
$m$ or $k$ can in principle be written as long as $p$ is large enough
\footnote{Pisarski and Rao\cite{PR} have argued that only in the Landau
gauge is the conventional propagator, Fig. 1a, b, c, infrared-finite at $p=0$,
although the $S$-matrix has only massive singularities.  Thus we expect our
$S$-matrix-derived propagator to have only massive singularities.}.  
We begin by looking at the small $m$, or equivalently large $p$ case,
which amounts to a perturbation expansion jointly in $Ng^2/p$ and in $m/p$.  We will save only the
lowest-order terms in these parameters.
Let us add to the bare inverse propagator (20) the sum of parity-even self-energies
in (26), (28), and (31) and parity-odd self-energies in (27), (29), and (32) at $m=0$ to find:
\begin{equation} \hat{\Delta}^{-1}(p)_{ij} = (p^2\delta_{ij}
- p_ip_j)(1 - 15Ng^2/32p) + \epsilon_{ija}p_ag^2(\frac{k + N}{4\pi})
\end{equation}
One recognizes here, as expected, the pure YM self-energy already given  
in (3) as the coefficient of $\delta_{ij}$.  In the $\epsilon$ term we
have rewritten $m$ of (20) as $kg^2/4\pi$, and we observe that there is a
$m^{-1}$ contribution to $\hat{B}$ at small $m$ which cancels out the
kinematic $m$ factor in (20), giving rise to the $N$ term in (33).  The result
is that $k$ is renormalized at one-loop level to $k + N$, a well-known\cite{PR}
result.  This is, in fact, the exact renormalization of $k$ to all orders of
perturbation theory\cite{PR}.  That the renormalization of $k$
is solely a mass renormalization can be traced to the QED-like Ward identity
of the type $Z_1=Z_2$ which holds for the gauge-invariant Green's functions
of the pinch technique.

Now we will look at small momentum for finite mass.  It is straightforward to check that $\hat{A}(p)$ is positive and
monotone decreasing (it vanishes at at rate $1/p$ at large $p$), so the
possibility of tachyonic behavior can be examined by looking at $p=0$.
At this point there is a perturbative expansion in powers of $Ng^2/m$.  The key result
comes from the parity-even self-energy at zero momentum:
\begin{equation} 1 - \hat{A}(p=0) = 1 - 29N/12k.\end{equation}
So if
\begin{equation} k \leq k_c \equiv 29N/12, \end{equation}
there is a tachyonic zero in the parity-even self-energy, or as pointed out
in equations (23), (25), a pole in the running mass and charge (unless
$\hat{B}$ were also to have a zero at the same $p$).

We conclude that one-loop perturbation theory fails for $k \leq k_c$.
Just because it does fail, we do not have a reliable value for the particular
value of $k$ where perturbation theory goes wrong.  For example, one might
argue that $k$ is renormalized to $k + N$ so that perhaps the critical value
of $k$ is found from $k + N \leq 29N/12$, or $k \leq 17N/12$.  This suggests
that we might be within a factor of two of understanding just what $k_c$ is;
to do better would require a fully non-perturbative treatment.  Although we
are not in a position to give this, we will in the next Section give some
qualitative arguments about what happens non-perturbatively when $k$ is
near $k_c$.

\section{Non-Perturbative Behavior Near the Critical Value of k}

Our objective here is to give a qualitative description of how YMCS theory
behaves as $k$ goes from small to large.  We will build on previous works
\cite{C89,C94} which have given a similar description for pure YM theory
(that is, $k=0$).  It was argued in these works that there is a condensate
of the operator $\theta = (G^a_{ij})^2/4$ with positive vacuum expectation
value, corresponding to a negative value of the vacuum energy $\epsilon$,
with $\epsilon = -\langle \theta \rangle /3$.  The basis for these arguments
is an exact form of the effective action as it depends on zero-momentum
matrix elements of $\theta$.  The condensate is
self-consistently related to the generation of a dynamical gauge-boson
mass $M$, with $M \sim (g^2 \langle\theta \rangle )^{1/4}$.  The negative
value of $\epsilon$ is associated with configurational entropy of the
condensate.  We first review these $k=0$ considerations, beginning with the definition of the gauge potentials and action for general $k$
via:
\begin{equation} \vec{A}(\vec{x}) = \frac{g\lambda_a}{2i}
\vec{A}^a(\vec{x}) \end{equation}
\begin{equation} G_{ij}(\vec{x}) = \partial_i A_j(\vec{x})
- \partial_j A_i(\vec{x}) + [A_i(\vec{x}), A_j(\vec{x})] \end{equation}
where the $\lambda_a$ are the usual generators of the group in the
fundamental representation.  The action is:
\begin{equation} S_{YMCS} = -\int d^3x\frac{1}{2g^2} Tr G_{ij}^2 +
2\pi ikW \equiv S_{YM} + 2\pi ikW \end{equation}
and the Chern-Simons term is
\begin{equation} W = -\frac{1}{8\pi^2}\int d^3x \epsilon_{ijk} Tr
(A_i\partial_j A_k - \frac{2}{3}A_iA_jA_k).\end{equation}
In these equations and in what follows a potential or field with a group index,
such as $\vec{A}^a$, is canonical and Hermitean, with $\vec{A}^a$ having mass
dimension 1/2; matrix-valued potentials and fields are multiplied by $g$ and are anti-Hermitean.  Note that $\theta$ is the square of the canonical field
strength and that $S_{YM} = \int d^3x \theta$.  The partition function and vacuum energy density are defined
as in (9) and (10), repeated here for convenience:
\begin{equation} Z = \int (dA) \exp(-S_{YMCS}) = \exp(-\int d^3x \epsilon)
\end{equation}

At $k=0$,the effective action\footnote{The vacuum energy $\int d^3x \epsilon$
is the effective action evaluated at its minimum in $\theta$.} $\Gamma(\theta)$ as it depends on the zero-momentum
matrix elements of $\theta$ is easily found\cite{C94} by introducing a
constant source $J$ into $Z$ and Legendre transforming:
\begin{equation} Z(J) \equiv \int (dA) \exp \{ \int d^3x\frac{1}{2g^2} Tr G_{ij}^2(1 - J)\}; \end{equation} 
that is, $S_{YM} = \int \theta$ so adding the constant source is just the same
as the change $g^2 \rightarrow g^2(1 - J)^{-1}$.  On dimensional grounds,
$\epsilon \sim g^6$ and thus
\begin{equation} \epsilon(J) = \epsilon(J=0)(1 - J)^{-3} =
-\frac{\langle \theta \rangle}{3}(1 - J)^{-3} \end{equation}
(by $\langle \theta \rangle$ we mean the expectation value of $\theta$ for $J=0$; see (43, 46) below).  Since $\theta = -\partial
\epsilon /\partial J$, one has
\begin{equation} \theta = \langle \theta \rangle (1 - J)^{-4}. \end{equation}  Legendre transform, as usual:
\begin{eqnarray} \Gamma(\theta) &  = & \epsilon(J) + \int d^3x J\theta\\
\partial \Gamma /\partial \theta & = & \int d^3x J. \nonumber \end{eqnarray}
One finds from (42-44) that
\begin{equation} \Gamma(\theta) = \int d^3x (\theta - \frac{4}{3}\theta^{3/4}
\langle \theta \rangle ^{1/4}), \end{equation}
with a minimum ($J=0$) value 
\begin{equation} \Gamma(\langle \theta \rangle) = -\frac{1}{3}\langle \theta \rangle \end{equation}

We have given earlier\cite{C89,C94} simple models which realize this
structure for $\Gamma(\theta)$, based on a one-dressed-loop CJT\cite{CJT}
potential where the dressed propagator has a dynamical mass $M$ depending 
on $\theta$.  The dressed propagator in question is the gauge-invariant
one called $\hat{\Delta}$ above.  Lavelle has shown\cite{La} that this
propagator has a gauge-invariant contribution\footnote{Non-gauge-invariant
condensates, such as ghosts, cancel out in $\hat{\Delta}$ but not in the
usual propagator.} from $\theta$ of the form
\begin{eqnarray} \hat{\Delta}_{ij}(p) & = & (\delta_{ij} - p_ip_j/p^2)
d^{-1}(p)\\
d(p) & = & p^2 + ag^2\theta/p^2 \end{eqnarray}
at large momentum $p$, where
\begin{equation} a = (\frac{58}{15})\frac{N}{N^2 - 1}. \end{equation}
The second term expresses the large-momentum behavior of the (squared)
dynamical mass $M$, just as a consituent quark mass is related to its
corresponding condensate.
At small momentum the $1/p^2$ behavior is modulated by the dynamical
mass $M$ (roughly, $p^2 \rightarrow p^2 + M^2$), as could be described in
principle by a non-linear Schwinger-Dyson equation (see Ref. \cite{C,CHK} for
an attempt in this direction which maintains gauge invariance).  Unfortunately, we do not really know how to do this, and we will instead make the crude
approximation of evaluating  the masslike term in (48) at $p^2=M^2$ to 
find $M$, or in other words,
\begin{equation} M = (ag^2\theta )^{1/4}. \end{equation}
A more accurate treatment would supply a constant factor of $O(1)$ in
this relation between mass and condensate.

The one-dressed-loop CJT approximation to the vacuum energy $\epsilon$
is really just a one-loop background field calculation\cite{H} with the
condensate field as the background field, which amounts to attaching external fields to the one-loop pinch-technique proper self-energy.  Instead of
calculating the self-energy perturbatively as we did in Section 1, we
give a mass $M$ by hand to the lines in this loop (for details, see Ref. \cite{C89,C94}), which gets rid of the perturbative infrared divergences
and mimics the result of solving the Schwinger-Dyson equations\cite{C,CHK}.
The result is
\begin{equation} \int d^3x \epsilon = \frac{V}{(2\pi )^3} \int d^3p
\frac{1}{4}\tilde{G}^a_{ij}(p)(1 - \hat{A}(p))\tilde{G}^a_{ij}(-p)
\end{equation}
where $V=\int d^3x, \tilde{G}$ is the Fourier-transformed field, and $\hat{A}(p)$ is the scalar integral for the pure YM theory pinch
self-energy but with massive propagators:
\begin{equation} \hat{A}(p) = (\frac{15}{4})Ng^2J_2 \end{equation}
(recall that $J_2$ is the $d=3$ scalar one-loop graph with two massive lines,
as given in (14) and (16)).  At $M=0$ this yields the massless pure-YM
result previously given in (3), (4), while at $p=0$ one finds
equation\begin{equation} 1 - \hat{A}(0) = 1 - 15Ng^2/32\pi M. \end{equation}
A simple qualitative approximation to the one-dressed-loop vacuum energy (51) comes from replacing $\hat{A}(p)$ by $\hat{A}(0)$ in that equation. One readily checks that, with the mass $M$ given by (50), the resulting one-dressed-loop
expression (51) for $\epsilon$ is of the desired form (45), with a specific
and probably rather inaccurate value for $\epsilon$ as a positive constant
times $(Ng^2)^3(N^2 -1)$.  The crucial minus sign in (45) comes from the
positive nature of $\hat{A}$ in (53), corresponding to asymptotic freedom.

Now turn to YMCS theory, with $k \neq 0$. 
It is easy to show that finite $k$ increases the vacuum energy, at a
quadratic rate for small $k$.
Note that for every (real) gauge potential $\vec{A}^a(\vec{x})$ occurring in
the path integral for $Z$ there is another configuration $-\vec{A}^a
(-\vec{x})$ for which the Yang-Mills part of the action, and the measure,
are unchanged, but for which the Chern-Simons term changes sign.  This
means that $Z$ is even in $k$, and we can write
\begin{equation} Z = \int (dA) \exp(-S_{YM}) \cos(2\pi kW) \end{equation}
and it is evident that $Z(k \neq 0) < Z(k=0)$.  A formal expansion for
small $k$ yields:
\begin{equation} \int d^3x \epsilon(k) = \int d^3x \epsilon(k=0)
+ 2\pi^2 k^2 \langle W^2 \rangle \end{equation}
where the expectation value is taken in the $k=0$ theory.
It is easy to check that $\langle W \rangle = O(k)$ for small $k$, so
to lowest order 
\begin{equation} \langle W^2 \rangle = \langle (W - \langle W \rangle )^2
\rangle \equiv \int d^3x \chi_{CS} \end{equation}
that is, the correction term depends on the gauge-invariant Chern-Simons
susceptibility $\chi_{CS}$, which has been studied in Ref. \cite{C94} for pure $SU(2)$ YM theory.  Standard $N$-counting arguments show that
\begin{equation} \epsilon \sim (N^2 - 1)(Ng^2)^3,\; \; \chi_{CS}
\sim (N^2 - 1)Ng^6 \end{equation}
so that the fractional increment in $\epsilon$ is $O(k^2/N^2)$.  It is not
hard to believe that there is some value of $k \sim N$ for which
$\epsilon =0$ and the condensate entropy no longer overcomes the free energy.
But for smaller values of $k,\; \epsilon$ remains negative and there is a
dominant condensate as well as a dynamical mass.  

We give a very simplified model of how that dynamical mass might interact
with the Chern-Simons mass.  The idea, as before,is that the non-linear Schwinger-Dyson equations for the YMCS pinch propagator will be singular if $m$ is too small, just as happens for the pure-YM theory, and the true mass of the theory will be augmented by a dynamical mass.  Actually to solve these Schwinger-Dyson equations is beyond our hopes at the moment, so we will proceed
with what we hope are sensible assumptions about what really happens and
fold in the dynamical mass $M$ more or less by hand into the $k=0$ results given above.   As described in Section 4 below, it is possible
to add a gauge-invariant mass term\cite{C74,C} to the YMCS action, yielding
an effective action describing the quantum-mechanical generation of dynamical
mass.  The resulting effective propagator is easy to describe in terms of the 
self-energies\footnote{The form given for $\hat{A}$, of opposite sign and with one more power of
$p$ than the one-loop perturbative result in (33), is characteristic of
what two- and higher-loop graphs contribute to mass generation.  It is what we expect from the condensate term in (48) at small momentum.}  $\hat{A},\;\hat{B}$ of Section 1 (equations (20-22)):  Just take
\begin{equation}1 - \hat{A}(p) = 1 + (M^2/p^2),\;\hat{B} = 1. \end{equation}
 One finds from
(22) that the parity-even propagator now has two Minkowski-space 
non-tachyonic masses $\mu_{\pm}$:
\begin{equation} \mu_{\pm} = \frac{1}{2}[\pm m + (m^2 + 4M^2)^{1/2}]
\end{equation}
which are also the physical masses coming from the Higgs mechanism\cite{PR}.
The mass $\mu_-$ goes to zero at $M=0$, and $\mu_+$ goes to $m$ in this 
limit.  We will make the {\it Ansatz}, unjustified by any deep analysis,
that a qualitative description of the behavior of the YMCS pinch propagator
at small momentum is found by replacing $m$ as it appears in the zero-momentum one-loop propagator (see (6), (7), (34))  by some mass which, like $\mu_+$, approaches $m$
as $M\rightarrow 0$.  (Recall that even though there is a zero-mass pole
in the bare YMCS propagator, it does not appear in physical quantities
like the $S$-matrix, so a mass like $\mu_-$ is not a good candidate.)  It 
is hardly justified to argue for the specific form of $\mu_+$, and instead
we use the replacement
\begin{equation} m \rightarrow \mu \equiv (m^2 + M^2)^{1/2} \end{equation}
which gives the correct small-$m$ or small-$k$ limit, and for large $m$
yields a mass which receives quadratic corrections from $M$.  This also holds
for $\mu_+$ of (59), but with a different numerical coefficient.

We want to investigate the behavior of the vacuum energy for $k \neq 0$
in the same spirit as before, in a one-dressed-loop approximation.  To
simplify matters we will only take explicit account of the parity-conserving
part of the one-loop pinch self-energy, which one can check does not change
the qualitative behavior we find.
Our strategy, then, is to take the $k=0$ expression (51) for the effective 
action and in it to make the replacement of 
\begin{equation} 1 - \hat{A}(p) \rightarrow 1 - \hat{A}(0) \approx
1 - 29Ng^2/48\pi \mu \end{equation}
where we used (7) for the dependence of $\hat{A}(0)$ on $m$.  Combine
equations (50, 51, 60, 61) to find:
\begin{equation} \epsilon = \theta \{1 - \frac{29Ng^2}{48 \pi}  
[m^2 + (ag^2\theta)^{1/2}]^{-1/2}\}. \end{equation}
At $m$ or $k=0$ this has the required pure-YM form (45), and for small
$k$ the fractional corrections to $\epsilon$ are also as required, to
increase it by $O(k^2/N^2)$.

Now let $m$ increase, and observe that when $k$ reaches the critical value
given earlier in equation (35), namely $29N/12$, the vacuum energy $\epsilon$
is positive for any positive value of $\theta$, and has its minimum at
$\theta = 0$.  The condensate and dynamical mass are now gone, and
semiclassical methods should serve to calculate all properties of YMCS
theory.  But for smaller $k$, $\epsilon$ has a minimum at some positive
value of $\theta$ and at the minimum $\epsilon$ is negative.  
To find that minimum, introduce the variables
\begin{equation} x = (ag^2\theta )^{1/2}/m^2,\;\;\alpha = 29N/12k. \end{equation}
The equation determining the value of $x$ for which $\epsilon$ is minimum is:
\begin{equation} (1 + x)^3 = \alpha^2(1 + \frac{3x}{4})^2.\end{equation}
At the critical value of $k$, which is $\alpha=1$, the only real 
solution is $x=0$.  For $k$ just below the critical value, or when
$\alpha - 1 \ll 1$, one finds $x \approx (4/3)(\alpha - 1)$ which
corresponds to the critical behavior
\begin{equation} M(\theta) \sim (\langle \theta \rangle )^{1/4}
\sim (k_c - k)^{1/2}.  \end{equation}
Finally, when $k \ll k_c$, the solution merges smoothly onto the $k=0$
model discussed above.  

We certainly cannot expect the crude approximations of this section to
be anywhere near quantitatively accurate, but we can hope that certain
features are realistic.  These might include the critical exponent $1/2$
for the dynamical mass given in (65).  It is unlikely that any real
improvement in the situation will be gained by analytic techniques, so
one might well look forward to lattice-gauge simulations of YMCS theory.

\section{Quantum Solitons}
D'Hoker and Vinet\cite{DV} have looked for Euclidean solitons of {\it classical}
YMCS theory.  It turns out that they found none with finite action, as we 
will review, either of the vortex type or of the sphaleron (i.e., spherically
symmetric) type.  Here we will investigate {\it quantum} solitons, that is,
solitons arising as solutions of an effective action, to be treated classically,
but containing an extra term summarizing the quantum-mechanical generation
of a dynamical mass $M$.  This effective action\cite{C,C74,CT,C89} is:
\begin{equation} S_{eff} = S_{YMCS} - M^2/g^2\int d^3x Tr(U^{-1}D_iU)^2
\end{equation}
where $D_i$ is the covariant derivative $\partial_i + A_i$ and $U$ is an
$N\times N$ matrix with the transformation law
\begin{equation} U \rightarrow VU \end{equation}
under local gauge transformations
\begin{equation} A_i \rightarrow VA_iV^{-1} + V\partial_iV^{-1} \end{equation}
The added mass term is just a gauged non-linear sigma model, and is locally
gauge-invariant under (67, 68).  The equations of motion are:
\begin{equation} [D_i,G_{ij}]  =  m^2(D_jU)U^{-1}\end{equation}
\begin{equation} [D_i,(D_iU)U^{-1}]  =  0   \end{equation}
Note that the equation of motion (70) for $U$ is compatible with the
identity
\begin{equation} [D_i, [D_j, G_{ij}]] \equiv 0 \end{equation}
As a result, not all the equations of motion are independent; those for 
$U$ follow from those for $A$.  We have already mentioned that the dynamical
mass $M$ is not really a constant, as we will treat it here, but vanishes
rapidly at large momentum (see (48)); in view of this rapid vanishing,
the mass term of the effective action (66) is not really correct at short
distance, and gives rise to spurious logarithmic divergences in the action
for the solitons we find below.  The true effective action is indeed finite,
and we will refer to our solitons as having finite action because the
YMCS part of the action is finite.    

Let us begin by finding vortex solutions with Abelian holonomy which, by
a local gauge transformation, can be chosen to be Abelian everywhere.
We give only solutions for closed vortices and not for open vortices which
terminate on monopoles.  These solutions are:
\begin{eqnarray} A_i(x) &  = & \frac{2\pi Q}{\mu}\oint dz_k \{ \epsilon_{ijk}\partial_j[\mu_-(\Delta_+(x-z)
-\Delta_0(x-z)) + (+ \leftrightarrow -)]\nonumber\\
& + & i\delta_{ik}\mu_+ \mu_-[\Delta_+ (x-z)
- \Delta_-(x-z)]\}
\end{eqnarray}
In this equation, the integral is over a closed path, and $Q$ is an anti-Hermitean generator of the gauge group such
that $\exp(2\pi Q)$ is in the center of the group\footnote{This normalization
is needed so that a gauge potential transported around a closed path linking
the vortex will be single-valued.}.  The quantities $\mu_{\pm}$ are the
masses given in equation (59), and $\mu = \mu_+ + \mu_-$.  The propagators
$\Delta_{\pm},\; \Delta_0$ are the usual Euclidean $d=3$ free propagators of
mass $\mu_{\pm}$ and 0, and the massless propagator gives the contribution
of the $U$ field.  This, of course, is the only surviving term at large
distances from the vortex closed path, and it is a singular pure gauge term.  The specific combination of propagators used is chosen
so that the vortex potential gives rise to finite YMCS action (per unit length).
Note that, in the $\epsilon_{ijk}$ term of (72) the combinations
$\Delta_{\pm} - \Delta_0$ and $\Delta_+ - \Delta_-$ are finite at short
distances.  

In the limit of pure YM theory( $\mu_+ = \mu_- = M$), the vortex solution
reduces to the previously-given\cite{C79,C95,CY} vortex which has only
the $\epsilon_{ijk}$ term in (72).  In the opposite limit $M=0,\; \mu_+ = m,\;
\mu_- = 0$, the vortex smoothly disappears.  It is apparent that the $\delta_{ik}$ term and the $\Delta_+$ part of the $\epsilon_{ijk}$ term disappear, because of the $\mu_-$ factor
(the propagators are not that singular in the limit).  The remaining 
$\epsilon_{ijk}$ term proportional to $\mu_+$ disappears because
$\Delta_- \rightarrow \Delta_0$ in the limit.  This is not to say that one
cannot find any vortex solution when $M=0$; in fact, one can, but it
involves only one mass $m$, not two, and the cancellations which occur in
(72) to yield finite action cannot be expressed.  The result is a vortex
with a singular field strength coming from the $\delta_{ik}$ term.

As long as $M \neq 0$, there is a long-range pure-gauge part of the vortex
(72) which can, in principle, contribute to confinement by a topological
linking of vortices with a Wilson loop\cite{C79,CY}.  But this can only
happen if there is in fact a condensate of vortices, and it is the non-zero condensate which is responsible for the existence of $M$.  So we expect
a string tension as long as there is a dynamical mass.  Based on earlier\cite{CY} work on the string tension in pure-YM theory, we might 
expect this quantity to scale like $M^2$, or in view of (65), like
$k_c - k$ near the transition point.

Note that the vortex is complex, with the real part even and the imaginary
part odd under $k \leftrightarrow -k$.  The action, including the Chern-
Simons term, is real, however.  

Next we turn to spherically-symmetric solitons, like the sphaleron\cite{MK,C77,DV,C89}.  We study explicitly only the $SU(2)$ case, and use the usual Witten\cite{W}
{\it Ansatz}, written in the form:
\begin{eqnarray}  2iA_i & = & \epsilon_{iak}\sigma_a \hat{x}_k
(\frac{\phi_1 - 1}{r})\nonumber\\& - & (\sigma_i - \hat{x}_i \hat{x}\cdot \sigma)
\frac{\phi_2}{r} + \hat{x}_i \hat{x}\cdot \sigma H_1 \end{eqnarray}  
\begin{equation} U = \exp[i(\beta /2)\hat{x} \cdot \sigma] \end{equation}
where $\hat{x}$ is the unit vector for $\vec{x}$ and all functions in (73-74)
depend only on the radial variable $r$.  As is well-known there is a
residual local $U(1)$ gauge invariance under which $\phi_1 + i\phi_2$
is multiplied by $\exp(i\lambda)$, $H_1$ transforms like a component of a $U(1)$ Abelian gauge potential, and $\beta \rightarrow \beta + \lambda(r)$.
We use this degree of freedom to choose $\beta = \pi$, which is the usual
choice\cite{MK,CT,C77,C89} for the $k=0$ sphaleron.  The equations of
motion\footnote{We correct some typographical sign mistakes in Refs.\cite{DV,C89} which did not appear in the equations these authors actually analyzed.  It should also be noted that the analysis of the small-$r$
behavior of $\phi_2$ and $H_1$ given in Ref. \cite{C89} is not sufficiently
general.}are (primes denote $d/dr$):
\begin{eqnarray} 0 & = & (\phi_1' - H_1\phi_2)' + \frac{1}{r^2}
\phi_1(1 - \phi_1^2 - \phi_2^2) \nonumber\\
&  & + (im - H_1)(\phi_2' + H_1\phi_1) - M^2(\phi_1 + 1) \end{eqnarray}
\begin{eqnarray} 0 & = & (\phi_2' + H_1\phi_1)' + \frac{1}{r^2}
\phi_2(1 - \phi_1^2 - \phi_2^2) \nonumber\\
& & - (im - H_1)(\phi_1' - H_1\phi_2) -M^2\phi_2 \end{eqnarray}
\begin{eqnarray} 0 & = & \phi_1\phi_2' - \phi_2\phi_1' + H_1(\phi_1^2
+ \phi_2^2) \nonumber\\
& & + \frac{im}{2}(1 - \phi_1^2 - \phi_2^2) + \frac{1}{2}M^2r^2H_1 \end{eqnarray}
\begin{equation} 0 = \frac{1}{2}(r^2H_1)' - \phi_2 \end{equation}
Equation (78) is the equation for $U$, and it is readily checked that it 
can be obtained by differentiating (77) and using (75-76).  We will use
(75-77) as a set of three independent equations.

The boundary conditions, for $\beta = \pi$, are:
\begin{equation} \phi_1(r=0) = 1;\;\phi_1(\infty ) = -1;\;
\phi_2(0) = \phi_2(\infty ) = 0. \end{equation}
The boundary conditions on $H_1$ follow from (77), which is an algebraic
equation for this quantity (see (81) below). Near $r=\infty $ the approach
to the values in (79) is exponential, while we find that $\phi_2$ approaches
zero linearly near $r=0$. 

These equations are complex, and have complex solutions, but with the
change $m \rightarrow im$ they would become real equations with real
solutions.  One can verify that it is consistent to choose $\phi_1$ to be
an even function of $m$ and the other functions to be odd.  Let us change to the dimensionless variables
\begin{equation} H_1 = imA(r),\; \phi_2 = i(m/M)B(r),\;x = Mr; \end{equation}
from now on, primes denote $d/dx$.
The equations of motion then become real, so we can choose $A$ and $B$ to be real also.  A further simplification of notation is to drop the
subscript on $\phi_1$, replacing it by $\phi$.  Just as for the vortex soliton, the action, including the Chern-Simons term, is real.

Note that we can solve for $H_1$ or $A$ algebraically (as could D'Hoker and
Vinet\cite{DV}, although their equations differ from ours):
\begin{equation} A = \frac{\phi 'B - B'\phi -(1/2)(1 - \phi^2 
+ (m/M)^2B^2)}{\phi^2 - (m/M)^2B^2 + (1/2)x^2} \end{equation}
D'Hoker and Vinet \cite{DV} have a similar equation, except that the
denominator is just $\phi^2$ (they use a gauge with $B=0$ and have no
mass $M$).  This creates singularities, since as we will see $\phi$ always
has at least one zero (exactly one for the $m=0$ sphaleron).  The feedback
of $A$ in the D'Hoker-Vinet case leads to infinitely many zeroes in $\phi$
with an accumulation point at the origin, and $\phi$ alternates between real and imaginary as it goes through its zeroes.  The D'Hoker-Vinet soliton has 
infinite action (a logarithmic singularity).  In our case, the mass term
in the denominator of (81) is potentially stabilizing, and the $B^2$ term
is potentially destabilizing.  However, for small $m/M$ this term is too small
to be harmful, since at small $x$, $\phi$ approaches unity.  So there is a
range of values of $m/M$ where the D'Hoker-Vinet singularity is cured and
finite-action solutions are expected.

We have studied these equations numerically, and find solutions for small
values of $m/M$.  As $m/M$ increases it becomes increasingly difficult to 
find a numerical solution, so we have tried a simple variational approach
with trial functions which are an excellent fit to the numerical solutions
where we can find them.  The variational approach indeed yields a minimum of
the action for small $m/M$ but for $m/M \gsim 0.5$ the denominator in the
equation (81) for $A$ is singular or nearly so at some point and we can go no farther.
Although we have no proof, it appears that there is no sphaleron-like
solution for larger values of $m/M$.
For small $m/M$ the sphaleron solution is just a perturbation of the
$m=0$ sphaleron\cite{C77,CT,C89}.  We display in Figs. 2, 3, 4 the
functions $\phi$, $A$, and $B$ as found from solving the unperturbed
equation for $\phi$ (the $m=0$ sphaleron), and the linearized equations
for $A,\;B$ in this field $\phi$.  The non-linear corrections to these
functions are $O(m^2/M^2)$.  Recall that the actual gauge potentials
differ from $A$ and $B$ by factors $\sim m$, as in (80), so $A$ and
$B$ go to zero linearly in this mass.

\section{Conclusions}
Although it is tempting to try, one cannot guess the fate of YMCS
theory as $k$ changes just from classical or semiclassical considerations.
Once loop effects are considered, in a gauge-invariant way, one begins to
see the same sort of tachyonic disease that is associated with pure $d=3$ YM theory (or with asymptotic
freedom in $d=4$), if $k \leq 29N/12$.  The cure for this disease is the generation of a dynamical mass $M$ through quantum effects.  The one-loop calculations we report in Section 2
are not quantitatively reliable, and this estimate of a critical
value of $k$ below which the theory is non-perturbative may well be off by a factor of two or so. This dynamical
mass $M$ is related in a self-consistent way to a condensate of quantum solitons which is supported
by configurational entropy.  These effects
may be crudely modeled by constructing a very approximate CJT one-dressed-loop
action which suggests the way in which the condensate and mass $M$ disappear
above the critical value of $k_c$.  Further support to these ideas comes
from Section 4, where the solitons of the theory are examined.  There are
finite-action solitons in the small-$k$ condensate regime, but these either
go away or become singular when the condensate disappears, supporting the
notion of self-consistency between the existence of solitons and of a 
condensate.

We do not know of any other studies of the issues addressed here, and so
there is nothing to compare with now.  Certainly it would be interesting to
make lattice-gauge simulations of YMCS theory, and we hope these are done.
In order to make sense out of the approach of $k$ to its critical value, one
would like to have $N$ large enough so that $k/N$ behaves somewhat like a
continuum parameter as $k$ changes by unit values, but this is not easy for lattice simulations.  

Let us recapitulate some of the estimates we have made:
\begin{enumerate}
\item For $k\leq k_c$, YMCS theory has a condensate, a dynamical mass, and
a string tension.  The critical value we estimate as
\begin{equation} k_c/N \approx 2 \pm .7;  \end{equation}
\item For $k$ near but less than $k_c$, the dynamical mass $M$ and condensate
$\langle \theta \rangle$ scale as:
\begin{equation} M \sim (k_c - k)^{1/2};  \end{equation}
\item Under the same conditions, the string tension $K_F$, proportional to
$M^2$, scales as:
\begin{equation} K_F \sim k_c - k .      \end{equation}
\end{enumerate}

It would, of course, be interesting to study further variants on pure YMCS
theory, e.g., adding fermions and scalars, and looking for non-perturbative
dynamics in supersymmetric versions.  Work in these directions is in progress.
\newpage
\section{Acknowledgements}
This work was supported in part by the National Science foundation under
grant PHY-9218990. 

\newpage
\begin{center}
{\large \bf Figure Captions}
\end{center}
Fig. 1.  One-loop $S$-matrix graphs from which the gauge-invariant
propagator is extracted.  Solid dots indicate a pinch vertex, as
described in the text.\\
\vspace{12pt}

\noindent Fig. 2.  The quantum soliton field $\phi$, for $m = 0$, plotted against
$x = Mr$.\\

\vspace{12pt}
\noindent Fig. 3.  The scaled quantum soliton field $B$ (defined in the text).
The actual field is found by multiplying by $m/M$ when this quantity is small
(higher-order corrections change $B$ by $O(m^2/M^2)$.\\

\vspace{12pt}
\noindent Fig. 4.  The scaled quantum soliton field $A$ (defined in the text).  The
actual field is found by multiplying by $m$, up to higher-order terms.
\newpage
\epsfig{file=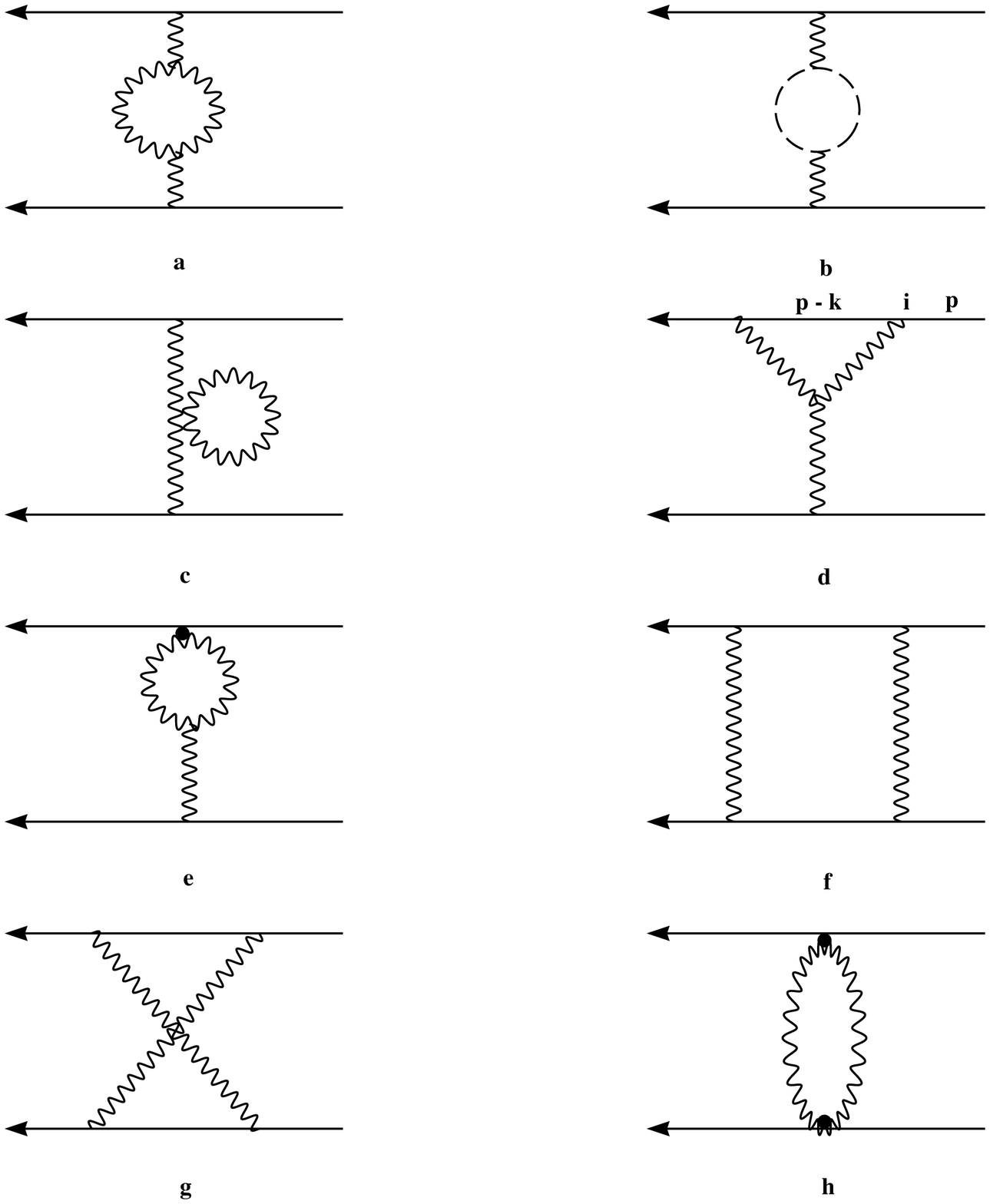,height=7in,clip=}
\epsfig{file=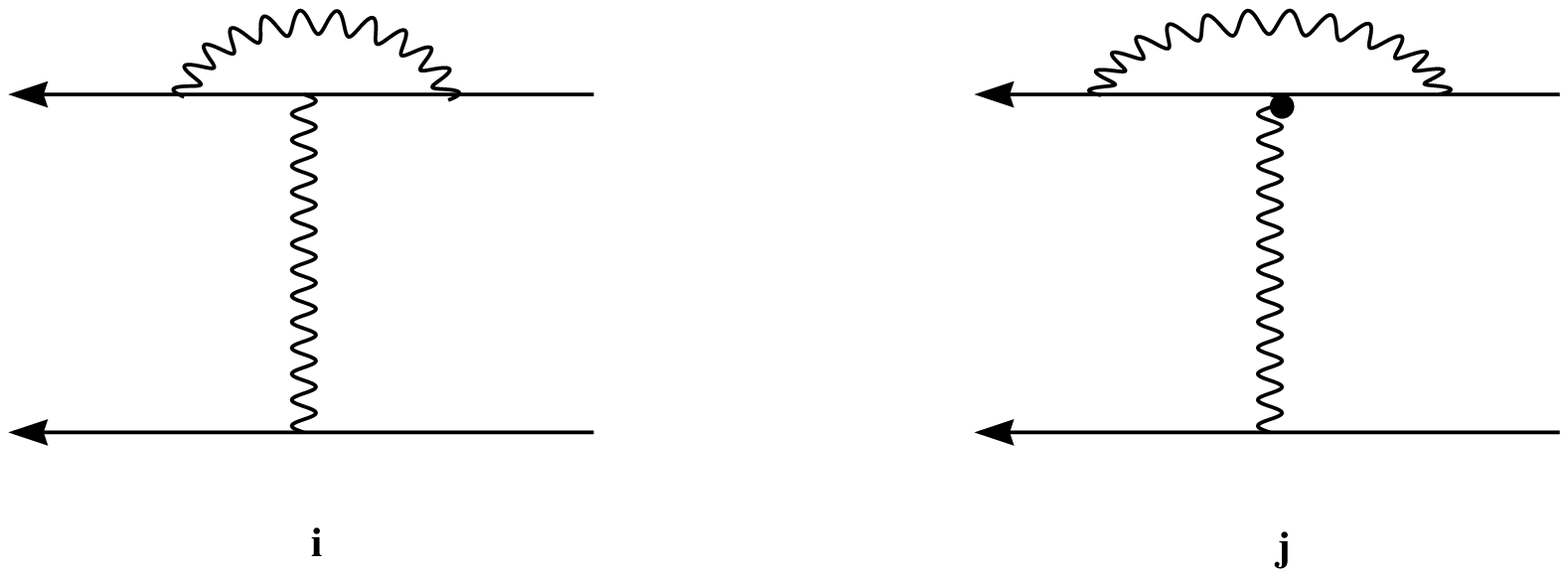,height=3in,clip=}
\epsfig{file=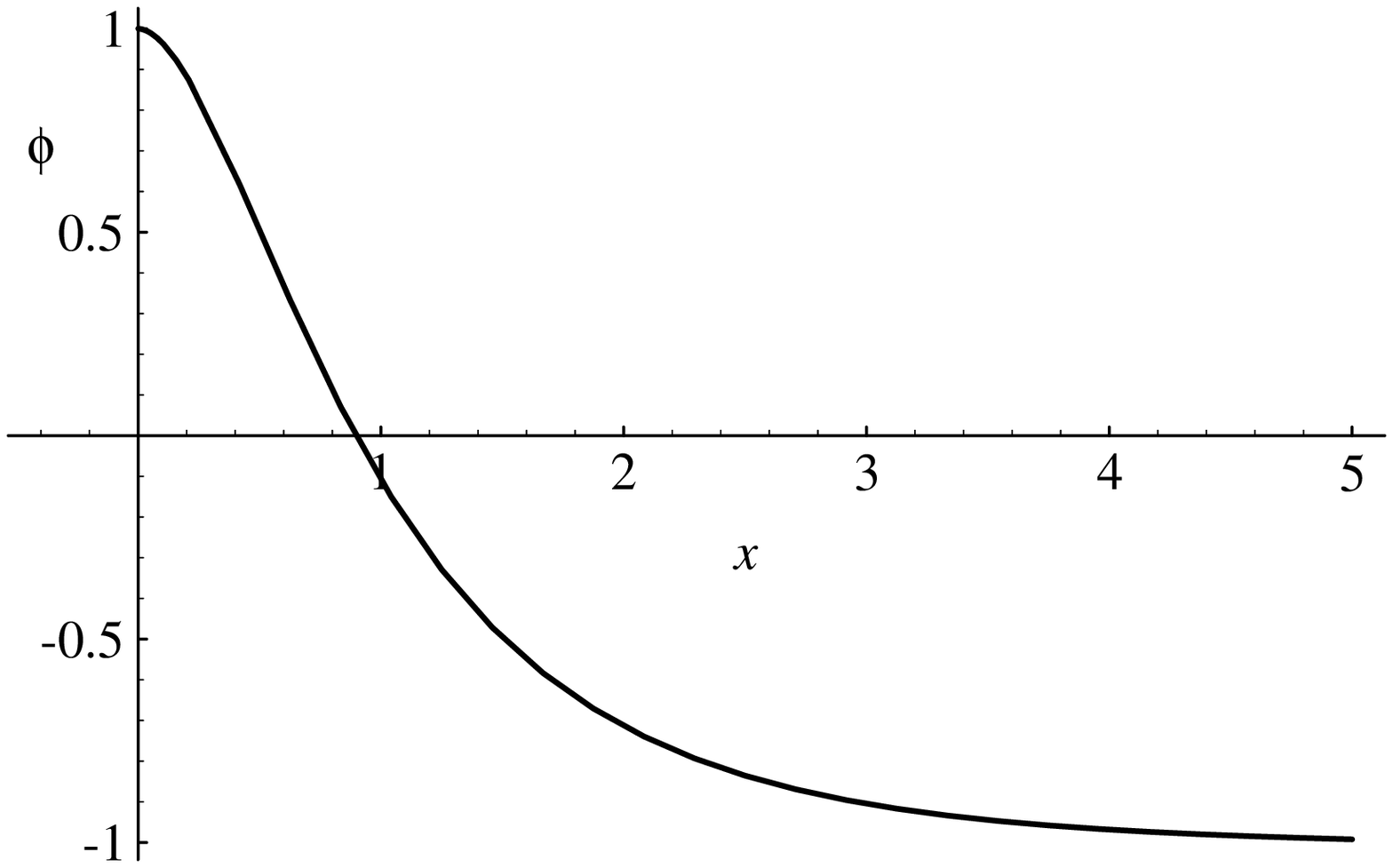,clip=}
\epsfig{file=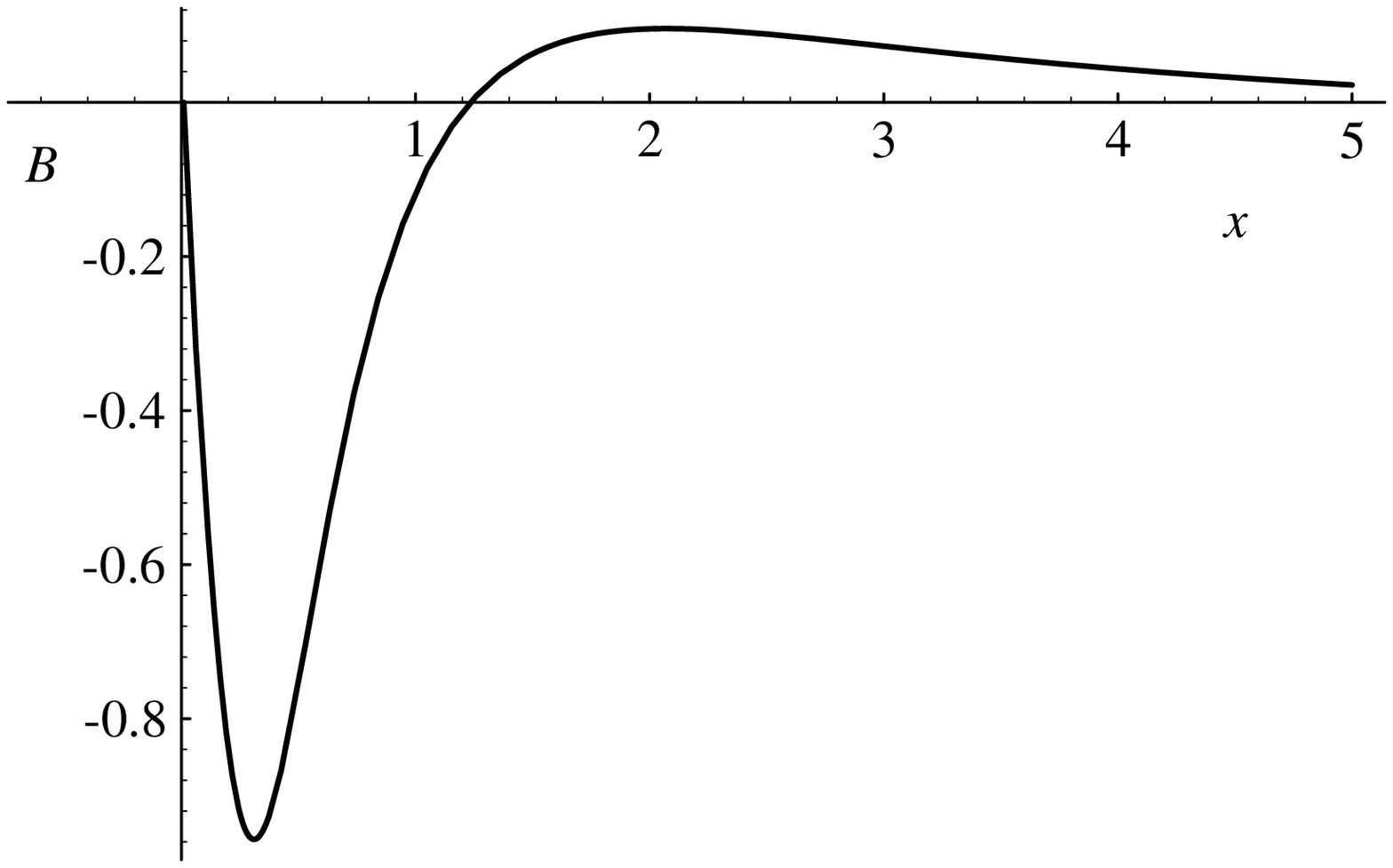,clip=}
\epsfig{file=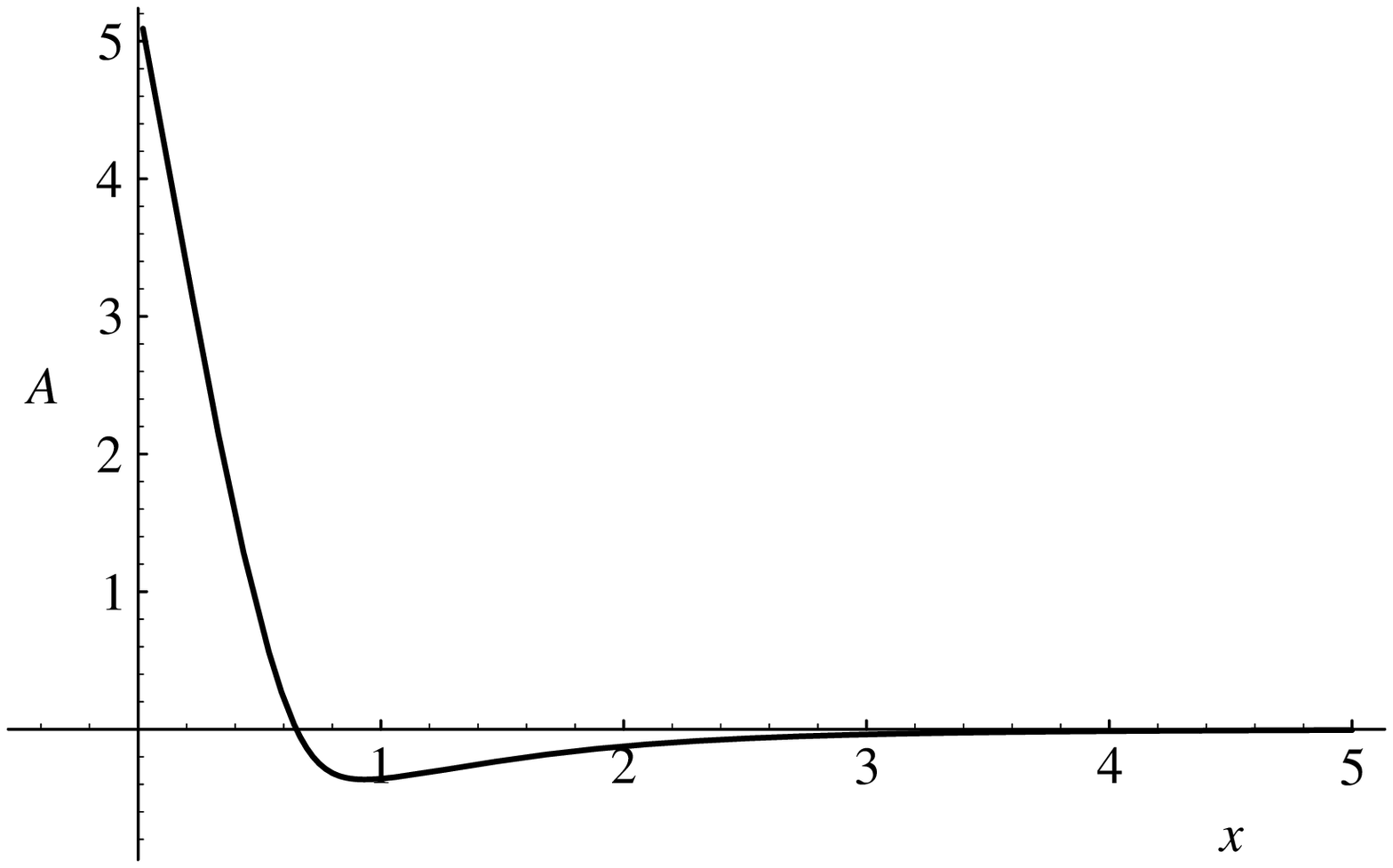,clip=}
\end{document}